\documentclass[aps,superscriptaddress,nofootinbib,eqsecnum,prd,notitlepage,twocolumn]{revtex4-1} 

\pdfoutput=1

\usepackage{amsfonts}
\usepackage{amsmath}
\usepackage{amssymb}
\usepackage{graphicx,color}
\usepackage{float}
\usepackage{hyperref}
\usepackage{subfigure}
\usepackage{dcolumn}
\usepackage{soul}
\usepackage{ulem}

\begin{document}

\title{Strengthening the de Sitter swampland conjecture in warm inflation}

\author{Robert Brandenberger}
\affiliation{Department of Physics, McGill University, Montreal, QC, H3A 2T8,
Canada}

\author{Vahid  Kamali}
\affiliation{Department of Physics, McGill University, Montreal, QC, H3A 2T8,
Canada}

\affiliation{Department of Physics, Bu-Ali Sina (Aviccena) University,
  Hamedan 65178, 016016, Iran}

\affiliation{School of Physics,
Institute for Research in Fundamental Sciences (IPM),
19538-33511, Tehran, Iran}

\author{Rudnei O. Ramos}
\affiliation{Departamento de Fisica Teorica, Universidade do Estado
  do Rio de Janeiro, 20550-013 Rio de Janeiro, RJ, Brazil }

\begin{abstract}

The {\it de Sitter constraint} on the space of effective scalar field
theories consistent with superstring theory provides a lower bound on
the slope of the potential of a scalar field which dominates the
evolution of the Universe, e.g., a hypothetical inflaton
field. Whereas models of single scalar field inflation with a
canonically normalized field do not obey this constraint, it has been
claimed recently in the literature that models of warm inflation can
be made compatible with it in  the case of large dissipation. The de
Sitter constraint is known to be derived from entropy considerations.
Since warm inflation necessary involves entropy production, it becomes
necessary to determine how this entropy production will affect the
constraints imposed by the swampland conditions. Here, we generalize
these entropy considerations to the case of warm inflation and show
that the condition on the slope of the potential remains essentially
unchanged and is, hence, robust even in the warm inflation
dynamics. We are then able to conclude that models of warm inflation
indeed can be made consistent with the {\it swampland} criteria.
 
\end{abstract}

\maketitle

\section{Introduction} 
\label{intro}

In order to understand the early evolution of the Universe at the
highest energy scales, a consistent embedding of the physical model of
the early Universe in a quantum theory of gravity is required. The
only currently known consistent quantum gravity theory which unifies
all forces of nature is superstring theory. Superstring theory leads
to tight constraints (the so-called {\it swampland constraints}) on
the space of possible effective field theories (EFT) which can be
applied to study the evolution of the Universe. In particular, scalar
fields arising in an EFT have a string-theoretical origin, and the
range of validity of such EFTs and the potentials for the scalar
fields are constrained. At the moment, the constraints are not at the
level of {\it theorems}, but at the level of {\it conjectures}.  The
first conjecture ({\it field range conjecture})~\cite{Ooguri:2006in}
states that the field range over which an effective field theory for a
canonically normalized scalar field $\phi$ is valid is bounded from
above by a constant of order one times $M_{\rm Pl}$, the four-dimensional (reduced)
Planck mass. The second condition ({\it de Sitter conjecture}) on the
EFT of a scalar field $\phi$ which is introduced to explain the
accelerated evolution of the early (or late)
Universe~\cite{Obied:2018sgi} is that its potential $V(\phi)$ has a
relative slope which is large in Planck units~\footnote{See, for
  instance Refs.~\cite{Vafa, Palti} for reviews, and
  Ref.~\cite{Samuel} for an understanding of the derivation of these
  conditions in the context of {\it String Gas
    Cosmology}~\cite{BV}.}. 

These conditions are opposite to the conditions required to obtain
successful cold~\footnote{We speak of {\it cold inflation} (CI) if the
  damping term in the scalar field equation of motion is dominated by
  Hubble damping.} slow-roll inflation, at least in models of single
field inflation based on a canonically normalized scalar field (the
{\it inflaton field})~\cite{Agrawal:2018own}, but they remain
consistent with quintessence models of Dark Energy~\cite{Lavinia}. On
the other hand, in {\it warm inflation} (WI)
models~\cite{Berera:1995ie} (for reviews, see, e.g.,
Refs.~\cite{Berera:2008ar,BasteroGil:2009ec}) both the distance and de
Sitter conjectures can be satisfied, as was discussed recently in
Refs.~\cite{Das:2018hqy, Motaharfar:2018zyb, Bastero-Gil:2018yen,
  Kamali:2019xnt, Berera:2019zdd, Kamali:2018ylz,
  Kamali:2019ppi,Kamali:2019hgv, Goswami:2019ehb} in the context of
high-dissipative models of WI. 

In Ref.~\cite{Ooguri:2018wrx}, a derivation of the de Sitter
conjecture was proposed which combines the distance criterion with an
entropy argument. It is postulated that the entropy within a Hubble
volume due to low mass bulk states cannot become larger than the
Gibbons-Hawking entropy~\cite{GH}. This is Bousso's~\cite{Bousso}
covariant entropy bound applied to a cosmological background. In the
case of WI, there is an extra source of bulk energy, namely the
thermal states into which part of the inflaton energy density is
converted to through dissipative effects. Dissipation effects during
slow-roll WI lead to a thermal bath of light particles which are
interacting with the
inflaton~\cite{Berera:2008ar,BasteroGil:2009ec}. In this work, we will
study the modification of the form of the de Sitter conjecture by
adding the extra entropy production during a period of WI. Since this
entropy production can potentially lead to a tighter constraint on the
slope of the potential, it needs to be carefully studied in order to
determine whether the Bousso bound can still remain satisfied. We
address this question and show that taking the additional bulk entropy
into account does not lead to an important change in the condition on
the slope of the potential and that, hence, WI remains compatible with
the swampland conjectures~\footnote{Note that  the {\it
Trans-Planckian Censorship Conjecture}, which has recently been
proposed~\cite{Bedroya1, Bedroya2} remains a challenge also for WI. 
This has also been addressed in ~\cite{Kamali:2019xnt,Berera:2019zdd}.}. The present
work then provides a demonstration of the robustness of WI  in
addressing the swampland conjectures in the context of entropy
considerations.

The paper is organized as follows. In Sec.~\ref{sec2}, we briefly
review the swampland conjectures and, in particular, the de Sitter
conjecture when addressed  through entropy considerations. In
Sec.~\ref{sec3} we summarize the WI dynamics and discuss its
application is the context of the present work. We explicitly show how
the entropy production in WI changes the Bousso condition. In
Sec.~\ref{sec4}, we discuss some general models of WI and different
primordial inflaton potentials demonstrating the robustness of WI as
far as its compatibility with the de Sitter conjecture is concerned.
{}Finally, in Sec.~\ref{concl}, we give our conclusions.

\section{Swampland conjectures through entropy considerations}
\label{sec2}

In Ref.~\cite{Ooguri:2006in} it was argued that the range of
applicability of any low energy EFT  which results after
compactification to four spacetime dimensions of a ten-dimensional
string theory is constrained. Specifically, the change $\triangle\phi$
of any scalar field arising in the EFT from its original point in
field space obeys the upper bound
\begin{eqnarray}
\frac{\triangle\phi}{M_{\rm Pl}} \, \leq \, c_1  ,
\label{distance}
\end{eqnarray} 
where $c_1$ is a constant of order one and  $M_{\rm Pl} \equiv
1/\sqrt{8 \pi G} \simeq 2.4 \times 10^{18}$GeV is the reduced Planck
mass, defined in terms of the usual Newton gravitational constant
$G$. Equation (\ref{distance}) is the {\it distance conjecture}. The
reason for this bound is that if the scalar field moves a distance
larger than that given by Eq.~(\ref{distance}), then a tower of new
string states becomes low mass and must be included in the low energy
EFT. 

Later~\cite{Obied:2018sgi, Garg, Ooguri:2018wrx}, it was argued that the self-interaction
potential of a scalar field which dominates the energy density of the
Universe is constrained by one of the following conditions -- either
\begin{eqnarray}
\frac{\mid\nabla V\mid}{V} \, \geq \, \frac{c_2}{M_{\rm Pl}},
\label{de Sitter1}
\end{eqnarray}
or
\begin{eqnarray}
\frac{min(\nabla_i\nabla_j V)}{V} \, \leq \, -\frac{c_3}{M_{\rm Pl}^2}
,
\label{de Sitter2}
\end{eqnarray}
where $\nabla$ is the gradient in field space, $c_2$ and $c_3$ are
universal and positive constants of order $1$ and
$min(\nabla_i\nabla_j V)$ is the minimum eigenvalue of the Hessian
$\nabla_i\nabla V_j$ in an orthonormal frame. This means that either
the potential is sufficiently steep, or else sufficiently
tachyonic. The condition (\ref{de Sitter2}) is used near the maximum
of the potential, if it has one. Note that these conditions on the
potential are trivially satisfied for a non-positive potential, or for
$M_{\rm Pl}\rightarrow\infty$ (see
Refs.~\cite{Ooguri:2006in,Agrawal:2018own,Ooguri:2018wrx}).

As shown in Ref.~\cite{Ooguri:2018wrx}, one can use entropy arguments
to derive the first part of de Sitter conjecture, given by Eq.~
(\ref{de Sitter1}), from the distance conjecture,
Eq.~(\ref{distance}), in the weak coupling regime of string theory,
where the physical observables are under control. To be specific, let
us assume that $\phi$ is an increasing function of time. The distance
conjecture states that as $\phi$ grows, there is an increasing number
$N(\phi)$ of effective degrees of freedom which can be excited and
which must enter into the low energy EFT. Specifically,
\begin{eqnarray}
N(\phi) \, \sim \, n(\phi)\exp\left(b\frac{\phi}{M_{\rm Pl}}\right),
\label{degree}
\end{eqnarray}
is the number of light particles with small masses,
\begin{eqnarray}
m \, \sim \, \exp(-a\triangle\phi) ,
\end{eqnarray}
where $a$ and $b$ in the above equations are constant parameters that
depend on the mass gap and other features of the tower of string
states and $n(\phi)$ is a monotonic function of $\phi$ with 
\begin{equation} 
\label{numberscaling}
\frac{dn(\phi)}{d\phi} \, > \, 0  .
\end{equation}

In an accelerating Universe with Hubble horizon $R=1/H$, the
increasing number of effective degrees of freedom  given by
Eq.~(\ref{degree})  leads to an increasing entropy of the tower of
states, which can be parametrized by
\begin{eqnarray}
S_{\rm tower}(N,R) \, = \, N^{\gamma}(M_{\rm Pl} R)^{\delta}  ,
\label{tower-entropy}
\end{eqnarray}
where $\gamma$ and $\delta$ are constants. {}For example, if the new
degrees of freedom behave as point particles, then $\delta = 0$. In
the accelerating Universe the entropy (\ref{tower-entropy}) is bounded
by the Gibbons-Hawking  entropy as~\cite{Bousso}
\begin{eqnarray} \label{ineq}
N^{\gamma}(M_{\rm Pl}R)^{\delta} \, \leq \, 8\pi^2 R^2M_{p}^2  .
\end{eqnarray}
Assuming that the potential energy dominates the total energy content,
we can express the Hubble rate in terms of the potential energy
density and rewrite Eq.~(\ref{ineq}) as
\begin{equation} 
\frac{V}{3M_{\rm Pl}^4} \, \leq \, \left( \frac{8 \pi^2}{N^{\gamma}}
\right)^{1/(1 - \delta/2)} .
\label{VN}
\end{equation}
{}By taking the logarithm of both sides of Eq.~(\ref{VN}) and then
differentiating with respect to $\phi$, we obtain
\begin{equation} 
\label{ineq2}
\frac{V'}{V} \, \leq \, - \frac{\gamma}{1 - \delta/2} ({\ln(N)})'  ,
\end{equation}
where the prime in the above equation and the following ones denotes a
derivative with respect to $\phi$ \footnote{Strictly speaking, this conclusion is valid provided that
we consider situations when the inequality (\ref{VN}) is close to being saturated.}.
With our convention for the running
of the field, $V'$ is negative and, hence, Eq.~(\ref{ineq2}) reads
\begin{equation}
\frac{|V'|}{V} \, \geq \, \frac{\gamma}{1 - \delta/2} (\ln(N))'
.
\end{equation}
Using Eqs.~(\ref{degree}) and (\ref{numberscaling}), we obtain
\begin{eqnarray}\label{p3}
\frac{|V'|}{V} & \geq & \frac{2\gamma}{2-\delta} \left(
\frac{n'}{n}+\frac{b}{M_{p}} \right)   \nonumber \\ & > &
\frac{1}{M_{p}} \frac{2b\gamma}{2-\delta}  ,
 \end{eqnarray}
where $\delta<2$ according to Ref.~\citep{Ooguri:2018wrx}.  The
result given by Eq.~(\ref{p3}) can be compared to the swampland de
Sitter condition Eq.~(\ref{de Sitter1}), where the constant $c_2$ is
then identified to be given by
\begin{equation}
c_2=\frac{2b\gamma}{2-\delta}.
\label{c2}
\end{equation} 

In the following section we will extend the above analysis to WI.

\section{Warm Inflation and de Sitter Conjecture}
\label{sec3}

In the WI scenario, dissipative effects intrinsic to the interactions
of the inflaton with other field degrees of
freedom~\cite{Berera:2008ar} can lead to a sustainable radiation
production throughout the inflationary expansion.  Those interactions
with other field degrees of freedom, for some region of parameters,
can be nonnegligible, generating dissipation terms and converting a
small fraction of vacuum energy density into radiation. When the
magnitude of these dissipation terms is strong enough to compensate
the redshift of the radiation by the expansion, a steady state can be
produced, with the inflationary phase happening in a thermalized
radiation bath. The background inflaton dynamics then becomes
\begin{equation}
\ddot \phi + ( 3 H + \Upsilon ) \dot \phi + V' =0,
\label{eomphi}
\end{equation}
while the evolution equation for the radiation fluid energy density
$\rho_r$ is
\begin{equation}
\dot \rho_r+ 4 H \rho_r  = \Upsilon \dot{\phi}^2, 
\label{eomrad}
\end{equation}
where $\Upsilon$ denotes the dissipative coefficient in WI and $H$ is
the Hubble parameter as usual. The entropy density $s$ is related to
the radiation energy density by $Ts=(1+w_r) \rho_r$, where $w_r$ is
the thermal bath equation of state.  Then,  Eq.~(\ref{eomrad}) can be
written in terms of the entropy density as
\begin{equation}
T \dot s+ 3 H Ts = \Upsilon \dot \phi^2 \,, 
\label{eoms}
\end{equation} 
where we have considered a thermalized radiation bath, i.e.,
$w_r=1/3$, as is typically the case in WI, where then 
\begin{equation}
\rho_r = \alpha  T^4,
\label{rhor}
\end{equation}
where  $\alpha =  \pi^2 g_*/30$ and $g_*$ is the effective number of
radiation degrees of freedom, while the associated entropy density is
\begin{equation}
s= \frac{4}{3} \alpha T^3,
\label{entropy}
\end{equation}

During the slow-roll epoch, Eqs.~(\ref{eomphi}), (\ref{eomrad}) and
(\ref{eoms}) become, respectively,
\begin{eqnarray}
&& 3H(1+Q)\dot{\phi}  \simeq  - V'  ,
\label{slowroll1}
\\ && \rho_r \simeq  \frac{3Q}{4} \dot{\phi}^2,
\label{slowroll2}
\\ && Ts \simeq Q \dot{\phi}^2,
\label{slowroll3}
\end{eqnarray}
where $Q$ denotes the dissipation rate in WI, defined as
\begin{equation}
Q \equiv \frac{\Upsilon}{3H},
\end{equation}
while for the Hubble parameter, since during WI we have that $V \gg
\rho_r$, we have
\begin{equation}
H^2 \simeq \frac{V}{3M_{\rm Pl}^2}.
\label{hubble}
\end{equation}

Combining Eqs.~(\ref{slowroll1}) and (\ref{slowroll2}) and also using
Eq.~(\ref{rhor}), we obtain
\begin{equation} 
\label{temprel}
\alpha T^4  \simeq \frac{(V')^2}{12 H^2} \frac{Q}{(1+Q)^2}.
\end{equation}

We can add the entropy of the thermal bath to the left-hand side of
Eq.~(\ref{ineq}). Demanding that the Bousso bound remains satisfied,
we are then led to the condition
\begin{eqnarray}\label{E2}
N^{\gamma}(M_{\rm Pl}R)^{\delta} + \frac{4}{3}\pi sR^3  \leq
8\pi^2(M_{\rm Pl}R)^2 ,
\end{eqnarray}
which implies that $N$ is bounded by
\begin{eqnarray}\label{E3}
N^{\gamma}M_{\rm Pl}^{\delta}  \leq  R^{3-\delta} \left( 8\pi^2
R^{-1}M_{\rm Pl}^2-\frac{4}{3}\pi s \right) .
\end{eqnarray}
Assuming that the potential energy density dominates the total energy
density, it then allows us to substitute for $R= 1/H$ making use of
the {}Friedmann equation in the slow-roll approximation,
Eq.~(\ref{hubble}), to obtain
\begin{equation}
R  =  \sqrt{3}\frac{M_{\rm Pl}}{\sqrt{V}} .
\end{equation}
Making use of Eq.~(\ref{degree}), taking the logarithm of both sides
of Eq.~(\ref{E3})  and differentiating with respect to the scalar
field, we obtain
\begin{eqnarray}\label{E7}
\gamma \left(  \frac{b}{ M_{\rm Pl}} + \frac{n'}{n} \right) & \leq &
-\frac{3-\delta}{2}\frac{V'}{V} \nonumber \\ & \times &  \left(
1-\frac{1}{3-\delta}\frac{1-\frac{\dot{s}}{6 \pi M_{\rm
      Pl}^2\dot{H}}}{1-\frac{s}{6\pi M_{\rm Pl}^2H}} \right)  ,
\end{eqnarray}
where we have made use of the relation
\begin{equation}
\frac{4\pi}{3} s'  \simeq  \frac{2\pi}{3} \frac{V'}{V} H
\frac{\dot{s}}{\dot{H}}\,.
\end{equation}

Comparing Eq.~(\ref{E7}) to Eq.~(\ref{p3}), we have that the WI
modification to the de Sitter swampland conjecture can be expressed as
\begin{equation}
\frac{|V'|}{V}  >  \tilde{c}_2,
\label{newdessitter}
\end{equation}
where $\tilde{c}_2$ is given by
\begin{eqnarray}
\tilde{c}_2 =  \frac{2\gamma b}{3-\delta}\left| 1-\frac{1}{3-\delta}
\frac{1-\frac{\dot{s}}{6\pi M_{\rm Pl}^2\dot{H}}} {1+\frac{s}{6\pi
    M_{\rm Pl}^2 H}}\right|^{-1}.
\label{c2tilde}
\end{eqnarray}  

To determine the magnitude of the WI modification in
Eq.~(\ref{c2tilde}), we need then to estimate values for the ratios
$s/H$ and $\dot s/\dot H$.  {}First, note that we can write the ratio
$s/H$ as
\begin{eqnarray}
\frac{s}{H}  &\simeq&  \frac{Q\dot{\phi}^2}{TH} \nonumber \\ &\simeq&
\frac{Q}{1+Q}2M_{\rm Pl}^2\frac{H}{T}\frac{\epsilon_{\phi}}{1+Q}
\nonumber \\ &\simeq& \frac{Q}{1+Q}2M_{\rm Pl}^2\frac{H}{T}\epsilon_H,
\label{s/H}
\end{eqnarray}
where we have used the slow-roll equations (\ref{slowroll1}),
(\ref{slowroll3}), (\ref{hubble}) and also that
\begin{equation}
\epsilon_\phi = \frac{M_{\rm Pl}^2}{2} \left(\frac{V'}{V} \right)^2,
\label{epsilonphi}
\end{equation}
and that in WI, 
\begin{equation}
\epsilon_H \equiv - \frac{\dot H}{H^2} = \frac{\epsilon_\phi}{1+Q}.
\label{epsilonH}
\end{equation}
Since in WI we have by definition that $T >H$, thus, we conclude that
the quantity $s/(6 \pi M_{\rm Pl}^2H)$  appearing in
Eq.~(\ref{c2tilde}) is much smaller than 1,
\begin{equation}
\frac{s}{6 \pi M_{\rm Pl}^2 H}  \simeq  \frac{1}{3 \pi}
\frac{Q}{1+Q}\frac{H}{T} \epsilon_H  \ll  1  ,
\label{sH2}
\end{equation}
which is valid in both the weak and strong dissipative regimes in WI,
i.e., $Q \ll 1$ and $Q \gg 1$,  respectively, and recalling that
current observations~\cite{Akrami:2018odb} limit the slow-roll
parameter to values $\epsilon_H \lesssim 10^{-2}$. We also note that
the result (\ref{sH2}) holds not only deep in the inflationary regime,
but it also remains true through the end of inflation, when
$\epsilon_H \to 1$. This is because typically the ratio $T/H$
increases towards the end of inflation, which is a common feature of
the known models of WI~\cite{Bastero-Gil:2016qru,Berera:2018tfc}.

Next, we notice that the term $\dot s/(6 \pi M_{\rm Pl}^2 \dot H)$ can
also be expressed as 
\begin{eqnarray}
\frac{\dot s}{6 \pi M_{\rm Pl}^2 \dot H} = \frac{s}{6 \pi M_{\rm Pl}^2
  H} \frac{s' H}{s H'},
\label{dotsH}
\end{eqnarray}
and
\begin{equation}
\frac{s' H}{s H'} \simeq 6 \frac{T' V}{T V'}.
\label{sHsH}
\end{equation}
Hence, Eq.~(\ref{c2tilde}) becomes,
\begin{eqnarray}
\tilde{c}_2 \simeq  \frac{2\gamma b}{3-\delta}  \Bigl| 1 &-&
\frac{1}{3-\delta} \left[ 1+ \kappa \left( 1 - 6 \frac{T' V}{T V'}
  \right)  \right.  \nonumber \\ &+& \left.  {\cal O }( \kappa^2 )
  \right]\Bigr|^{-1},
\label{c2tilde2}
\end{eqnarray}  
where
\begin{equation}
\kappa = \frac{s}{6 \pi M_{\rm Pl}^2 H} .
\end{equation}

To study the change in the coefficient $\tilde{c}_2$ between the cases
of CI and WI  we can parametrize the dissipation coefficient
$\Upsilon$ as follows,
\begin{equation}
\Upsilon(T,\phi) = C T^n \phi^p M^{1-n-p},
\label{Upsilon}
\end{equation}
where $C$ is a dimensionless constant (that carries the details of the
microscopic model used to derive the dissipation coefficient, e.g.,
the different coupling constants of the model), $M$ is a mass scale in
the model and depends of its construction, while  $n$ and $p$ are
numerical powers, which can be either positive or negative
numbers. The parametrization Eq.~(\ref{Upsilon}) covers many different
models of WI that have been proposed in the literature (see, e.g.,
Ref.~\cite{Kamali:2019xnt} and references therein). As an additional
note, we should also point out that the dynamical  stability of the
background equations in WI requires that the power $n$ in the
dissipation coefficient Eq.~(\ref{Upsilon}) satisfies $-4 < n < 4$,
which is valid in the weak ($Q \ll 1$) and strong ($Q>1$) dissipative
regimes of WI~\cite{Moss:2008yb,delCampo:2010by,BasteroGil:2012zr}. 

{}From Eqs.~(\ref{temprel}) and (\ref{Upsilon}) we find that the
functional relation between the temperature and the amplitude of the
inflaton field can be expressed, in the high dissipative regime $Q>1$,
as
\begin{eqnarray}
T(\phi)  \simeq \left[ \frac{\sqrt{3}  M_{\rm Pl}}{4 \alpha} \frac{
    (V')^2}{C \phi^p V^{1/2} M^{1-n-p} } \right]^{\frac{1}{4+n}},
\label{Tphi1}
\end{eqnarray}
and, consequently, we find
\begin{eqnarray}
\frac{T'}{T}  = \frac{1}{4+n} \left(
2\frac{V''}{V'}-\frac{1}{2}\frac{V'}{V}-\frac{p}{\phi} \right) .
\label{TT}
\end{eqnarray}
Substituting Eq.~(\ref{TT}) in the ratio $T' V/(T V')$ appearing in
Eq.~(\ref{c2tilde2}), we find that
\begin{eqnarray}
\frac{T' V}{T V'}\Bigr|_{Q>1} \simeq  \frac{1}{4+n} \left[ 2\frac{V
    V''}{(V')^2}- \frac{p V}{\phi V'} -\frac{1}{2} \right].
\label{c2tildeQlarge}
\end{eqnarray}  
Likewise, working now in the low dissipative regime of WI, $Q \ll 1$,
we find that
\begin{eqnarray}
T(\phi)  \simeq \left[ \frac{\sqrt{3}  M_{\rm Pl}^3}{12 \alpha} \frac{
    C \phi^p M^{1-n-p} (V')^2} {V^{3/2} } \right]^{\frac{1}{4-n}},
\label{Tphi2}
\end{eqnarray}
and
\begin{eqnarray}
\frac{T'}{T}  = \frac{1}{4-n} \left(
2\frac{V''}{V'}-\frac{3}{2}\frac{V'}{V}+\frac{p}{\phi} \right),
\label{TT2}
\end{eqnarray}
which gives for the ratio $T' V/(T V')$ the result
\begin{eqnarray}
\frac{T' V}{T V'}\Bigr|_{Q \ll 1} \simeq \frac{1}{4-n} \left[ 2\frac{V
    V''}{(V')^2} +  \frac{p V}{\phi V'} -\frac{3}{2} \right].
\label{c2tildeQsmall}
\end{eqnarray}

\section{Examples}
\label{sec4}

Using the results given by Eqs.~(\ref{c2tildeQlarge}) and
(\ref{c2tildeQsmall}), we can find the form of $\tilde{c}_2$  once the
primordial inflaton potential is given. Let us consider some
representative models next as examples. 

\subsection{Monomials chaotic inflation models}

Let us consider the case of monomial chaotic inflation models given
by the primordial potential,
\begin{equation}
V(\phi) = \frac{V_0}{2 k} \left(\frac{\phi}{M_{\rm Pl}} \right)^{2k}.
\label{monomial}
\end{equation}
Cases with $k=1,\,2$ and $3$ are known to satisfy the observational
constraints in the WI
scenario~\cite{Benetti:2016jhf,Bastero-Gil:2019gao}.

Using Eq.~(\ref{monomial}) in Eqs.~(\ref{c2tildeQlarge}) and
(\ref{c2tildeQsmall}), we find, respectively,
\begin{equation}
\frac{T' V}{T V'}\Bigr|_{Q>1}  = \frac{1}{4+n} \left( \frac{3}{2} -
\frac{p+2}{2k} \right),
\label{mon1}
\end{equation}
and
\begin{equation}
\frac{T' V}{T V'}\Bigr|_{Q \ll 1} = \frac{1}{4-n} \left( \frac{1}{2} +
\frac{p-2}{2k} \right).
\label{mon2}
\end{equation}

We can consider in the above equations two previous well studied
dissipation coefficients in WI and with a quartic inflaton potential
($k=2$): (a) A cubic in the temperature dissipation
coefficient~\cite{Bartrum:2013fia}, $\Upsilon = C T^3/\phi^2$, for
which we have $n=3$ and $p=-2$. This gives for Eq.~(\ref{mon1}) the
result $3/14$, while for Eq.~(\ref{mon2}) it gives $-1/2$; (b) A
linear in the temperature dissipation
coefficient~\cite{Bastero-Gil:2016qru}, $\Upsilon = C T$, for which we
have $n=1$ and $p=0$. This gives for Eq.~(\ref{mon1}) the result
$1/5$, while for Eq.~(\ref{mon2}) it gives $0$.  This remains true
also for other more general dissipation coefficients, like the ones
recently studied in
Refs.~\cite{Bastero-Gil:2019gao,Berghaus:2019whh}. In all these cases
we find that the ratio $T' V/(T V')$ is in general smaller than 1.
Thus, $\tilde{c}_2 \simeq c_2$ is obtained for all these WI models.

\subsection{Hilltop type of inflation models}

Hilltop type of inflaton models are described by a potential of the
form,
\begin{equation}
V(\phi) = V_0 \left[ 1- \beta \left(\frac{\phi}{M_{\rm Pl}}
  \right)^{2k} \right].
\label{hilltop}
\end{equation}
These type of potentials, representing small field like models of
inflations, can also be shown to be compatible with the observations
in WI~\cite{Benetti:2016jhf}, with the constant $\beta$ in
Eq.~(\ref{hilltop}) in general satisfying $\beta \sim 10^{-2}$.

As in the previous example, we find now that
\begin{eqnarray}
&&\frac{T' V}{T V'}\Bigr|_{Q>1} 
\nonumber \\ 
&&= \frac{1}{4+n} \left\{ \frac{ \left[ p
    - 2 (2k-1) \right]   \left[ 1-\beta \left(\frac{\phi}{M_{\rm Pl}}
    \right)^{2k} \right]}{2 \beta k \left(\frac{\phi}{M_{\rm Pl}}
  \right)^{2k}} - \frac{1}{2} \right\},
\nonumber \\
\label{mon3}
\end{eqnarray}
and
\begin{eqnarray}
&& \frac{T' V}{T V'}\Bigr|_{Q \ll 1} 
\nonumber \\
&&= -\frac{1}{4-n} \left\{
\frac{\left[ p + 2 (2k-1) \right]   \left[ 1-\beta
    \left(\frac{\phi}{M_{\rm Pl}} \right)^{2k} \right]}{2 \beta k
  \left(\frac{\phi}{M_{\rm Pl}} \right)^{2k}} + \frac{3}{2}  \right\}.
\nonumber \\
\label{mon4}
\end{eqnarray}

In the example studied for instance in Ref.~\cite{Benetti:2016jhf},
where $k=1$ and $\beta \sim 10^{-2}$, we typically have $\phi > 4
M_{\rm Pl}$ and ultraviolet completions for the potential
(\ref{hilltop}) are also expected in general to be in a scale $m_{\rm
  UV} \gg M_{\rm Pl}$ when the potential is no longer
unbounded. Considering also the most common dissipation terms used in
the literature for WI, we expect that both Eqs.~(\ref{mon3}) and
(\ref{mon4}) to be not larger than order 1.  Thus, here also we can
predict that $\tilde{c}_2 \simeq c_2$.

\subsection{Generalized exponential type of inflation models}

Inflation models of a generalized exponential form given by
\begin{equation}
V(\phi) = V_0 e^{ -\beta \left(\frac{\phi}{M_{\rm Pl}} \right)^{k} },
\label{exppot}
\end{equation}
have also been studied recently in the literature and found regions of
parameter space  compatible with the observations in the WI scenario,
at least for cases with $k>2$ (see, e.g., Ref.~\cite{Lima:2019yyv}). 

As in the previous examples, we now have that
\begin{equation}
\frac{T' V}{T V'}\Bigr|_{Q>1}  = \frac{1}{4+n} \left[
  \frac{p-2(k-1)}{\beta k \left(\frac{\phi}{M_{\rm Pl}} \right)^{k} }
  +\frac{3}{2}   \right],
\label{mon5}
\end{equation}
and
\begin{equation}
\frac{T' V}{T V'}\Bigr|_{Q \ll 1} = -\frac{1}{4-n} \left[
  \frac{p+2(k-1)}{\beta k \left(\frac{\phi}{M_{\rm Pl}} \right)^{k} }
  -\frac{1}{2}   \right].
\label{mon6}
\end{equation}

In the type of inflation models with the potential (\ref{exppot}) a
scaling regime is fast approached when $\beta (\phi /M_{\rm Pl})^k
\gtrsim 1$ after the inflationary regime.  In this case, both
Eqs.~(\ref{mon3}) and (\ref{mon4}) are found to be not larger than
order 1.  Even during inflation, at least for the examples studied in
Ref.~\cite{Lima:2019yyv}, we find that the product $\kappa [1-6 T'
  V/(T V')]$ in Eq.~(\ref{c2tilde2}) is always found to be much
smaller than 1. Thus, here we also find  $\tilde{c}_2 \simeq c_2$.

\section{Conclusions}
\label{concl}

We have generalized the derivation of the {\it de Sitter constraint}
on effective field theories consistent with quantum gravity to the
case of WI. We have shown that the condition on the slope of the
scalar field potential is changed in a minimal way due to the entropy
effects inherent of the WI dynamics.  We have studied general classes
of dissipation terms currently found in the WI literature and which
are able to lead to consistent observational quantities. Different
broad classes of primordial inflaton potentials were also
studied. {}For all these cases, we found that the WI correction to the
de Sitter  swampland condition is always negligible. Hence, the {\it
  warm inflation} scenario can indeed be made compatible with the {\it
  swampland conditions} and we have demonstrated here that this is a
robust result.

\section*{Acknowledgments}

The research at McGill is supported in part by funds from NSERC and
from the Canada Research Chair program. The visit of VK to McGill has
also been supported in part by the McGill Space Institute.  R.O.R. is
partially supported by research grants from Conselho Nacional de
Desenvolvimento Cient\'{\i}fico e Tecnol\'ogico (CNPq), Grant
No. 302545/2017-4, and Funda\c{c}\~ao Carlos Chagas Filho de Amparo
\`a Pesquisa do Estado do Rio de Janeiro (FAPERJ), Grant
No. E-26/202.892/2017.



\begin{thebibliography}{99}


\bibitem{Ooguri:2006in}  H.~Ooguri and C.~Vafa, ``On the Geometry of
  the String Landscape and the Swampland,'' Nucl.\ Phys.\ B {\bf 766},
  21 (2007) doi:10.1016/j.nuclphysb.2006.10.033 [hep-th/0605264].
  
\bibitem{Obied:2018sgi} 
  G.~Obied, H.~Ooguri, L.~Spodyneiko and C.~Vafa,
  ``De Sitter Space and the Swampland,''
  arXiv:1806.08362 [hep-th].
  
 
 
   
\bibitem{Vafa}
T.~D.~Brennan, F.~Carta and C.~Vafa,
 ``The String Landscape, the Swampland, and the Missing Corner,''
 PoS TASI {\bf 2017}, 015 (2017)
 doi:10.22323/1.305.0015
 [arXiv:1711.00864 [hep-th]].
 
\bibitem{Palti}
 E.~Palti,
 ``The Swampland: Introduction and Review,''
 arXiv:1903.06239 [hep-th].
 
\bibitem{Samuel}
 S.~Laliberte and R.~Brandenberger,
  ``String Gases and the Swampland,''
  arXiv:1911.00199 [hep-th].
   
\bibitem{BV}      
 R.~H.~Brandenberger and C.~Vafa,
 ``Superstrings In The Early Universe,'' Nucl.\ Phys.\ B {\bf 316}, 391 (1989).
  
\bibitem{Agrawal:2018own} 
  P.~Agrawal, G.~Obied, P.~J.~Steinhardt and C.~Vafa,
  ``On the Cosmological Implications of the String Swampland,''
  Phys.\ Lett.\ B {\bf 784}, 271 (2018)
  doi:10.1016/j.physletb.2018.07.040
  [arXiv:1806.09718 [hep-th]].
 
\bibitem{Lavinia}
L.~Heisenberg, M.~Bartelmann, R.~Brandenberger and A.~Refregier,
  ``Dark Energy in the Swampland,''
  Phys.\ Rev.\ D {\bf 98}, no. 12, 123502 (2018)
  doi:10.1103/PhysRevD.98.123502
  [arXiv:1808.02877 [astro-ph.CO]].
    
\bibitem{Berera:1995ie} 
  A.~Berera,
  ``Warm inflation,''
  Phys.\ Rev.\ Lett.\  {\bf 75}, 3218 (1995)
  doi:10.1103/PhysRevLett.75.3218
  [astro-ph/9509049].
  
\bibitem{Berera:2008ar} 
  A.~Berera, I.~G.~Moss and R.~O.~Ramos,
  ``Warm Inflation and its Microphysical Basis,''
  Rept.\ Prog.\ Phys.\  {\bf 72}, 026901 (2009)
  doi:10.1088/0034-4885/72/2/026901
  [arXiv:0808.1855 [hep-ph]].

\bibitem{BasteroGil:2009ec} 
  M.~Bastero-Gil and A.~Berera,
  ``Warm inflation model building,''
  Int.\ J.\ Mod.\ Phys.\ A {\bf 24}, 2207 (2009)
  doi:10.1142/S0217751X09044206
  [arXiv:0902.0521 [hep-ph]].
 
\bibitem{Das:2018hqy} 
  S.~Das,
  ``Note on single-field inflation and the swampland criteria,''
  Phys.\ Rev.\ D {\bf 99}, no. 8, 083510 (2019)
  doi:10.1103/PhysRevD.99.083510
  [arXiv:1809.03962 [hep-th]].
  
\bibitem{Motaharfar:2018zyb} 
  M.~Motaharfar, V.~Kamali and R.~O.~Ramos,
  ``Warm inflation as a way out of the swampland,''
  Phys.\ Rev.\ D {\bf 99}, no. 6, 063513 (2019)
  doi:10.1103/PhysRevD.99.063513
  [arXiv:1810.02816 [astro-ph.CO]].
  
\bibitem{Bastero-Gil:2018yen} 
  M.~Bastero-Gil, A.~Berera, R.~Hern\'andez-Jim\'enez and J.~G.~Rosa,
  ``Warm inflation within a supersymmetric distributed mass model,''
  Phys.\ Rev.\ D {\bf 99}, no. 10, 103520 (2019)
  doi:10.1103/PhysRevD.99.103520
  [arXiv:1812.07296 [hep-ph]].
  
\bibitem{Kamali:2019xnt} 
  V.~Kamali, M.~Motaharfar and R.~O.~Ramos,
  ``Warm brane inflation with an exponential potential: a consistent realization away from the swampland,''
  Phys.\ Rev.\ D {\bf 101}, no. 2, 023535 (2020)
  doi:10.1103/PhysRevD.101.023535
  [arXiv:1910.06796 [gr-qc]].
  
\bibitem{Berera:2019zdd} 
  A.~Berera and J.~R.~Calder\'on,
  ``Trans-Planckian censorship and other swampland bothers addressed in warm inflation,''
  Phys.\ Rev.\ D {\bf 100}, no. 12, 123530 (2019)
  doi:10.1103/PhysRevD.100.123530
  [arXiv:1910.10516 [hep-ph]].
 
\bibitem{Kamali:2018ylz} 
  V.~Kamali,
  ``Non-minimal Higgs inflation in the context of warm scenario in the light of Planck data,''
  Eur.\ Phys.\ J.\ C {\bf 78}, no. 11, 975 (2018)
  doi:10.1140/epjc/s10052-018-6449-x
  [arXiv:1811.10905 [gr-qc]].
 
\bibitem{Kamali:2019ppi} 
  V.~Kamali,
  ``Warm pseudoscalar inflation,''
  Phys.\ Rev.\ D {\bf 100}, no. 4, 043520 (2019)
  doi:10.1103/PhysRevD.100.043520
  [arXiv:1901.01897 [gr-qc]].
 
\bibitem{Kamali:2019hgv} 
  V.~Kamali,
  ``Reheating After Swampland Conjecture,''
  JHEP {\bf 2001}, 092 (2020)
  doi:10.1007/JHEP01(2020)092
  [arXiv:1902.00701 [gr-qc]].
 
 


\bibitem{Goswami:2019ehb} 
  G.~Goswami and C.~Krishnan,
  ``Swampland, Axions and Minimal Warm Inflation,''
  arXiv:1911.00323 [hep-th].
  
  
  
  
  
  
  
  
  
 \bibitem{Ooguri:2018wrx} 
  H.~Ooguri, E.~Palti, G.~Shiu and C.~Vafa,
  ``Distance and de Sitter Conjectures on the Swampland,''
  Phys.\ Lett.\ B {\bf 788}, 180 (2019)
  doi:10.1016/j.physletb.2018.11.018
  [arXiv:1810.05506 [hep-th]].
 
    
\bibitem{GH}
G.~W.~Gibbons and S.~W.~Hawking,
  ``Cosmological Event Horizons, Thermodynamics, and Particle Creation,''
  Phys.\ Rev.\ D {\bf 15}, 2738 (1977).
  doi:10.1103/PhysRevD.15.2738
  
\bibitem{Bousso} 
  R.~Bousso,
  ``A Covariant entropy conjecture,''
  JHEP {\bf 9907}, 004 (1999)
  doi:10.1088/1126-6708/1999/07/004
  [hep-th/9905177].
 
\bibitem{Bedroya1}
A.~Bedroya and C.~Vafa,
  ``Trans-Planckian Censorship and the Swampland,''
  arXiv:1909.11063 [hep-th].
  
\bibitem{Bedroya2}
 A.~Bedroya, R.~Brandenberger, M.~Loverde and C.~Vafa,
  ``Trans-Planckian Censorship and Inflationary Cosmology,''
  arXiv:1909.11106 [hep-th].
  
  
 \bibitem{Garg} 
  S.~K.~Garg and C.~Krishnan,
  ``Bounds on Slow Roll and the de Sitter Swampland,''
  JHEP {\bf 1911}, 075 (2019)
  doi:10.1007/JHEP11(2019)075
  [arXiv:1807.05193 [hep-th]].
  
\bibitem{Akrami:2018odb} 
  Y.~Akrami {\it et al.} (Planck Collaboration),
  ``Planck 2018 results. X. Constraints on inflation,''
  arXiv:1807.06211.
  
\bibitem{Bastero-Gil:2016qru} 
  M.~Bastero-Gil, A.~Berera, R.~O.~Ramos and J.~G.~Rosa,
  ``Warm Little Inflaton,''
  Phys.\ Rev.\ Lett.\  {\bf 117}, no. 15, 151301 (2016)
  doi:10.1103/PhysRevLett.117.151301
  [arXiv:1604.08838 [hep-ph]].
  
\bibitem{Berera:2018tfc} 
  A.~Berera, J.~Mabillard, M.~Pieroni and R.~O.~Ramos,
  ``Identifying Universality in Warm Inflation,''
  JCAP {\bf 1807}, 021 (2018)
  doi:10.1088/1475-7516/2018/07/021
  [arXiv:1803.04982 [astro-ph.CO]].
  
\bibitem{Moss:2008yb} 
  I.~G.~Moss and C.~Xiong,
  ``On the consistency of warm inflation,''
  JCAP {\bf 0811}, 023 (2008)
  doi:10.1088/1475-7516/2008/11/023
  [arXiv:0808.0261 [astro-ph]].
  
\bibitem{delCampo:2010by} 
  S.~del Campo, R.~Herrera, D.~Pav\'on and J.~R.~Villanueva,
  ``On the consistency of warm inflation in the presence of viscosity,''
  JCAP {\bf 1008}, 002 (2010)
  doi:10.1088/1475-7516/2010/08/002
  [arXiv:1007.0103 [astro-ph.CO]].

\bibitem{BasteroGil:2012zr} 
  M.~Bastero-Gil, A.~Berera, R.~Cerezo, R.~O.~Ramos and G.~S.~Vicente,
  ``Stability analysis for the background equations for inflation with dissipation and in a viscous radiation bath,''
  JCAP {\bf 1211}, 042 (2012)
  doi:10.1088/1475-7516/2012/11/042
  [arXiv:1209.0712 [astro-ph.CO]].
  
\bibitem{Benetti:2016jhf} 
  M.~Benetti and R.~O.~Ramos,
  ``Warm inflation dissipative effects: predictions and constraints from the Planck data,''
  Phys.\ Rev.\ D {\bf 95}, no. 2, 023517 (2017)
  doi:10.1103/PhysRevD.95.023517
  [arXiv:1610.08758 [astro-ph.CO]].
  
\bibitem{Bastero-Gil:2019gao} 
  M.~Bastero-Gil, A.~Berera, R.~O.~Ramos and J.~G.~Rosa,
  ``Towards a reliable effective field theory of inflation,''
  arXiv:1907.13410 [hep-ph].
  
\bibitem{Bartrum:2013fia} 
  S.~Bartrum, M.~Bastero-Gil, A.~Berera, R.~Cerezo, R.~O.~Ramos and J.~G.~Rosa,
  ``The importance of being warm (during inflation),''
  Phys.\ Lett.\ B {\bf 732}, 116 (2014)
  doi:10.1016/j.physletb.2014.03.029
  [arXiv:1307.5868 [hep-ph]].
  

\bibitem{Berghaus:2019whh} 
  K.~V.~Berghaus, P.~W.~Graham and D.~E.~Kaplan,
  ``Minimal Warm Inflation,''
  arXiv:1910.07525 [hep-ph].

\bibitem{Lima:2019yyv} 
  G.~B.~F.~Lima and R.~O.~Ramos,
  ``Unified early and late Universe cosmology through dissipative effects in steep quintessential inflation potential models,''
  Phys.\ Rev.\ D {\bf 100}, no. 12, 123529 (2019)
  doi:10.1103/PhysRevD.100.123529
  [arXiv:1910.05185 [astro-ph.CO]].
    
\end{thebibliography}
\end{document}